\documentclass[manuscript]{aastex}

\usepackage{amsbsy,amssymb,amsmath,bm}
\usepackage{graphicx,subfigure}
\usepackage{natbib}
\usepackage{color}

\begin{document}

\title{Estimations and scaling laws for stellar magnetic fields}
\author{Xing Wei}
\affil{Department of Astronomy, Beijing Normal University, China \\ Email: xingwei@bnu.edu.cn}

\begin{abstract}
In rapidly rotating turbulence (Rossby number much less than unity), the standard mixing length theory for turbulent convection breaks and Coriolis force enters the force balance such that magnetic field eventually depends on rotation. By simplifying the self-sustained magnetohydrodynamics dynamo equations of electrically conducting fluid motion, with the aid of theory of isotropic non-rotating or anisotropic rotating turbulence driven by thermal convection, we make estimations and derive scaling laws for stellar magnetic fields with slow and fast rotation. Our scaling laws are in good agreement with the observations.
\end{abstract}
\maketitle

\section{Motivation}\label{sec:motivation}

\begin{figure}
\centering
\includegraphics[scale=1.0]{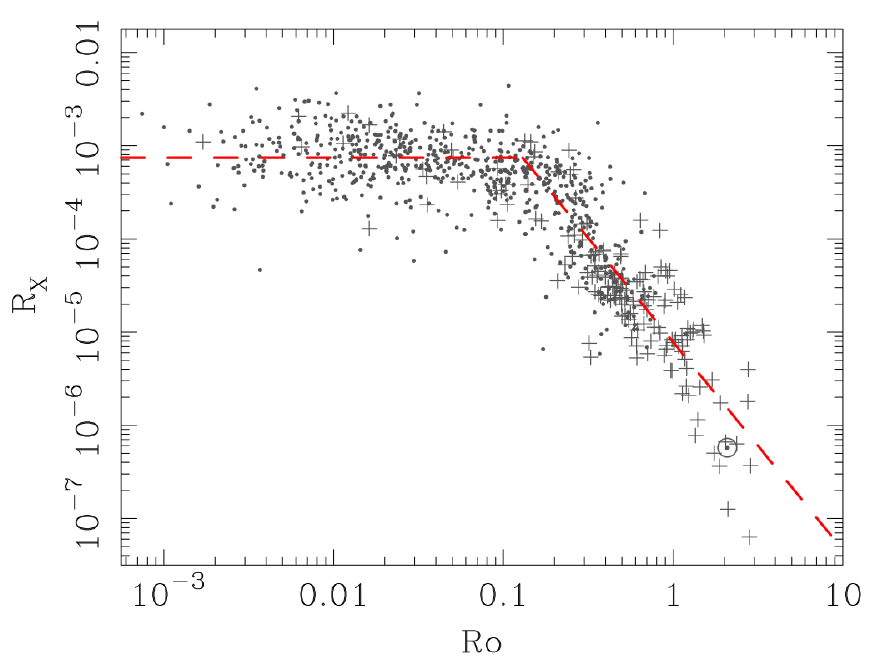}
\caption{Figure in \citet{wright2011}. Horizontal axis denotes Rossby number and vertical axis denotes the fraction of stellar X ray luminosity. The red line shows the fitting.}\label{fig}
\end{figure}

The observations show that the fraction of stellar X ray luminosity $L_X/L_{bol}$ increases with stellar rotation rate until it reaches saturation for sufficiently fast rotation \citep{wright2011, vidotto2014, reiners2014}. Figure \ref{fig} extracted from \citet{wright2011} shows this relation between $L_X/L_{bol}$ and rotation. Rossby number $Ro=P_{rot}/\tau$, the ratio of stellar rotation period to turbulent convection timescale, measures rotation. \citet{wright2011} gives the scaling law $L_X/L_{bol}\propto Ro^{-2}$ for the stage before saturation that field increases with rotation. The X ray emission is caused by mass loss near stellar surface, which arises from surface magnetic field with open field lines. Surface field stems from internal field which is generated through dynamo in convection zone. i.e. magnetic field is amplified by shear and twist of field lines due to differential rotation and turbulent convection of electrically conducting fluid. Therefore, the fraction $L_X/L_{bol}$ represents the strength of stellar magnetic field \citep{Pevtsov2003, vidotto2014}.

To interpret the observations, the interface $\alpha$-$\Omega$ dynamo model \citep{Montesinos2001} and the flux transport model are proposed in the observational papers mentioned above. However, both models describe kinematic dynamo, i.e. fluid motion is prescribed but not coupled with magnetic field (Lorentz force or back reaction of field on flow is neglected), and cannot interpret the self-sustained turbulent dynamo in stellar convection zone. Moreover, the mean-field dynamo was widely used to interpret the field-rotation relation. For example, \citet{blackman2015} used the kinematic $\alpha$-$\Omega$ mean-field dynamo model, \citet{blackman2016} used the dynamic mean-field dynamo model in which rotation, magnetic field and mass loss are coupled, and \citet{Kitchatinov2015} used the Babcock-Leighton model which is also an $\alpha$-$\Omega$ kinematic mean-field dynamo model. In addition to mean-field dynamo, magnetic helicity was introduced together with mean-field model to interpret astrophysical dynamos \citep{Vishniac2001, Vishniac2014} or corona activity \citep{blackman2000} and ejection \citep{blackman2003}. However, the mean-field dynamo is a parameterized model which solves only magnetic induction equation in the absence of fluid dynamics equation, namely it does not solve the full magnetohydrodynamics (MHD) equations, especially in the lack of thermal convection. The flow is usually characterized by a differential rotation plus the parameter $\alpha$ which arises from helical motion \citep{moffatt1978}. Although in \citet{blackman2003} the back reaction of magnetic field on the parameter $\alpha$ was considered, namely Lorentz force enters the expression of $\alpha$, and in \citet{blackman2016} the effect of Lorentz force on differential rotation was considered, both did not solve the full MHD equation, especially thermal convection.

In this paper, we study the convection driven dynamo by simplifying the self-sustained MHD equations, with the aid of the turbulence theory for thermal convection with or without rotation, to make estimations and derive scaling laws for magnetic energy at both slow and fast rotation rates, and then compare our predictions with the observations. We focus on the range in which magnetic field depends on rotation ($Ro>0.1$ in Figure \ref{fig}) and give a tentative interpretation about the saturation range ($Ro<0.1$ in Figure \ref{fig}).

\section{Estimation and scaling laws}

How planetary magnetic fields depend on physical properties (e.g. density, radius, rotation, etc.) is extensively studied, and a good summary can be found in the review paper \citet{christensen2010}. In \citet{christensen2009} a scaling law for magnetic fields of rotating planets and stars is proposed with the aid of the standard mixing length theory for isotropic non-rotating turbulence. Although this scaling law is for rotating planet or star, it is independent of rotation itself, because the energy equation which was used to derive this scaling does not involve rotation, i.e. Coriolis force does not enter energy equation (we will see this later). However, as we have already shown in the last section, observations confirm that stellar magnetic fields depend on rotation. We will illustrate the reason in this section, i.e. the standard mixing length theory breaks for rapidly rotating turbulence. Here fast or slow rotation is evaluated with Rossby number, i.e. whether it is greater or less than unity. For planetary field, the mixing length cannot exceed the depth of convection zone, so that \citet{christensen2009} can be applied to planetary field, but it cannot well interpret the dependence of stellar field on rotation. In \citet{davidson2013} another scaling law for magnetic fields of rotating planets, which depends on rotation, is proposed with the aid of theory for anisotropic rotating turbulence. But it is only for planetary fields. We now study stellar magnetic fields.

We start from the magnetohydrodynamics equations of electrically conducting fluid motion. The momentum equation in a frame rotating at angular velocity $\bm\Omega$ reads
\begin{equation}\label{ns}
\rho\left(\frac{\partial\bm v}{\partial t}+\bm v\cdot\bm\nabla\bm v\right)=-\bm\nabla p+2\rho\bm v\times\bm\Omega+\bm J\times\bm B+\delta\rho\,\bm g+\rho\bm f_\nu
\end{equation}
and the magnetic induction equation reads
\begin{equation}\label{induction}
\frac{\partial\bm B}{\partial t}=\bm\nabla\times(\bm v\times\bm B)+\eta\nabla^2\bm B.
\end{equation}
The variables are used with their conventional notation and $\delta\rho$ is density deviation from surrounding because of thermal convection. On the right-hand-side of \eqref{ns} the terms are pressure force, Coriolis force, Lorentz force, buoyancy force and viscous force. Performing $\bm v\cdot$ \eqref{ns} plus $\bm B\cdot$ \eqref{induction} we obtain the energy equation
\begin{equation}\label{energy}
\frac{\partial}{\partial t}\left(\frac{\rho v^2}{2}+\frac{B^2}{2\mu}\right)=-\bm\nabla\cdot\bm A+\delta\rho\,\bm g\cdot\bm v-D_\nu-\frac{J^2}{\sigma}.
\end{equation}
The left-hand-side is the rate of total energy. On the right-hand-side, vector $\bm A$ is the total flux, consisting of kinetic energy flux $(\rho v^2/2)\bm v$, pressure energy flux $p\bm v$ and Poynting flux $\bm E\times\bm B/\mu$. The second term is the power of buoyancy force. The third term $D_\nu$ is viscous dissipation. The last term is Ohmic dissipation, where $\sigma=1/(\mu\eta)$ is electric conductivity. We take the volume average for \eqref{energy} in the convection zone, i.e. $(1/V)\int\eqref{energy}dV$. When dynamo saturates the total energy is statistically steady so that the left-hand-side vanishes. By divergence theorem $\bm\nabla\cdot\bm A$ vanishes because the net flux across the surface is very small, namely the energy loss from the surface due to stellar wind and Poynting flux is negligible compared to buoyancy energy in the interior. Although mass loss due to stellar wind or magnetic helicity may be noticeable \citep{blackman2003}, we think that the magnetic energy that mass loss carries away is a tiny fraction of the magnetic energy generated by convection dynamo in the interior. Viscous dissipation is very small compared to Ohmic dissipation. Thus, the two terms are left to balance each other
\begin{equation}\label{balance}
\langle\delta\rho\,gv\rangle\approx\langle\frac{J^2}{\sigma}\rangle
\end{equation}
where bracket denotes volume average. Equation \eqref{balance} states that the power of buoyancy force is almost equal to Ohmic dissipation rate, which is exactly the essence of convection dynamo. It should be noted that Coriolis force due to rotation does not enter the energy equation, since it is perpendicular to fluid velocity.

Next we introduce two lengthscales, the mixing length $l$ for turbulent momentum transfer, i.e. the size of the largest turbulent eddies, and the lengthscale $l_B$ of magnetic field. The mixing length is usually assumed to be twice of pressure scale height 
\begin{equation}\label{l}
l\approx\frac{2p}{dp/dr}\approx\frac{2p}{\rho g}\approx\frac{2\mathcal{R}T}{\mu_m g}
\end{equation}
where hydrostatic balance $dp/dr\approx\rho g$ and equation of state for ideal gas $p=(\mathcal{R}/\mu_m)\rho T$ are employed ($\mathcal{R}$ is gas constant and $\mu_m$ is mean molecular weight). To obtain the estimation for $l_B$ we return back to \eqref{induction}. The three terms are estimated as follows. Magnetic field temporally varies on the convective timescale $l/v$, i.e. $\partial\bm B/\partial t\sim vB/l$. The magnetic induction term takes effect on the large lengthscale $l$, i.e. $\bm\nabla\times(\bm v\times\bm B)\sim vB/l$ which is comparable to the time-derivative term. Magnetic diffusion term takes effect on the small lengthscale $l_B$, i.e. $\eta\nabla^2\bm B\sim \eta B/l_B^2$. This two-scale analysis with respect to $l$ and $l_B$ is widely used in dissipative systems (e.g. our system with magnetic diffusion or Ohmic dissipation). The three terms are on the comparable order of magnitude, and this balance immediately yields the estimation
\begin{equation}\label{l_B}
\frac{l_B}{l}\approx\left(\frac{vl}{\eta}\right)^{-1/2}=Rm^{-1/2}
\end{equation}
where magnetic Reynolds number is defined on the large lengthscale $l$, i.e. $Rm=vl/\eta$. Ampere's law $\bm\nabla\times\bm B=\mu\bm J$ yields the estimation $J\approx B/(\mu l_B)$. Inserting $J\approx B/(\mu l_B)$ and \eqref{l_B} into \eqref{balance}, we obtain
\begin{equation}\label{magnetic-buoyancy}
\langle\frac{B^2}{\mu}\rangle\approx\langle\delta\rho\,gl\rangle
\end{equation}
which states the equipartition between magnetic energy and buoyancy energy. Equation \eqref{magnetic-buoyancy} is what we will use to estimate magnetic energy in the next.

We study two cases, one is slow rotation with $Ro>1$ and the other is fast rotation with $Ro<1$. With slow rotation we adopt the standard mixing length theory
\begin{equation}\label{buoyancy-kinetic}
\delta\rho\,gl\approx\rho v^2
\end{equation}
which states the equipartition between buoyancy energy and kinetic energy. Combing \eqref{magnetic-buoyancy} and \eqref{buoyancy-kinetic}, we find that the three energies (magnetic, buoyancy, kinetic) are at the same order of magnitude. If we change the view from energy to force, \eqref{buoyancy-kinetic} indicates the balance between buoyancy force and inertial force, i.e. $\delta\rho\,g\approx \rho v^2/l$. With slow rotation, this force balance can be applied. However, with fast rotation, inertial force is negligible compared to Coriolis force. Then the force balance is built between buoyancy force and Coriolis force, i.e. $\delta\rho\,g\approx \rho v_\perp\Omega$ where $v_\perp$ is velocity perpendicular to rotation axis (the parallel component has no contribution to Coriolis force). In the next we will see that this force balance in rapidly rotating turbulence is numerically validated. Moreover, the lengthscale in the estimation \eqref{magnetic-buoyancy} will be the eddy lengthscale parallel to rotational axis $l_\parallel$, and we will also address this point in the next.

With slow rotation, we combine \eqref{magnetic-buoyancy} and \eqref{buoyancy-kinetic} to find $\langle B^2/\mu\rangle\approx\langle\rho v^2\rangle$. Now we need to estimate convective velocity $v$. Instead of $\delta\rho$ which cannot be observed, we use heat flux $F$ which is related to luminosity (observable quantity) to measure convection
\begin{equation}\label{F}
F=\rho c_p\delta T\,v=c_pT\delta\rho\,v
\end{equation}
where thermodynamics relation $\delta T/T=-\delta\rho/\rho$ for ideal gas at constant pressure is employed (what we are concerned with is magnitude so that we omit the minus sign). Then \eqref{l}, \eqref{buoyancy-kinetic} and \eqref{F} combine to yield the estimation
\begin{equation}\label{v}
v\approx\left(\frac{F}{\rho}\right)^{1/3}
\end{equation}
where $c_p=2.5\mathcal{R}/\mu_m$ in convection zone is employed. The estimation for convective velocity \eqref{v} is already validated by numerical simulations \citep{chan1996, cai2014}. Inserting \eqref{v} into $\langle B^2/\mu\rangle\approx\langle\rho v^2\rangle$, we obtain the estimation for magnetic energy with slow rotation
\begin{equation}\label{B-slow}
\langle\frac{B^2}{\mu}\rangle\approx\langle\rho^{1/3}F^{2/3}\rangle
\end{equation}
which is similar to \citet{christensen2009}, i.e. magnetic energy is independent of rotation rate. More strictly speaking, \citet{christensen2009} chose the lengthscale in \eqref{magnetic-buoyancy} to be the minimum of the mixing length \eqref{l} and the depth of convection zone. For most planets with small size, this lengthscale will be the depth of convection zone, but for most stars it will be the mixing length. Therefore, for a rapidly rotating planet, this scaling law independent of rotation is fine because the depth of convection zone does not depend on rotation. However, for a rapidly rotating star, this lengthscale will be $l_\parallel$ which strongly depends on rotation (we will see how strong this dependence is in the next). That is the reason why \citet{christensen2009} cannot well interpret the stellar field-rotation relation \citep{wright2011}.

With fast rotation, turbulence is anisotropic and has a large-scale columnar structure along rotation axis. The formation of such two-dimensional turbulence structure is caused by propagation of inertial waves at its group velocity (inertial waves are induced by Coriolis force). Suppose that an eddy with the initial size $l$, namely the mixing length in the absence of rotation, is elongated at the group velocity of inertial waves $c_g\approx\Omega l$ until the eddy turnover timescale $\tau\approx l/v_\perp$ (i.e. the lifetime of an eddy with size $l$) where $v_\perp$ is the turbulent velocity perpendicular to rotation axis. This eddy will grow until its length parallel to rotation axis reaches $l_\parallel\approx c_g\tau\approx\Omega l^2/v_\perp$ when the eddy is destroyed in turbulence at its lifetime $\tau$. The most recent numerical simulations of rotating turbulent convection find that turbulent velocity is suppressed by fast rotation \citep{cai2021}, i.e. the velocity parallel to rotation axis scales as $v_\parallel\approx vRo^{5/3}$ and the velocity perpendicular to rotation axis scales as $v_\perp\approx vRo^{7/3}$ ($v_\perp$ is even more suppressed than $v_\parallel$), where $v$ is the turbulent convective velocity in the absence of rotation and Rossby number is defined as $Ro=v/(\Omega l)$. We return back to $l_\parallel$ to readily find the scaling law $l_\parallel\approx\Omega l^2/v_\perp\approx lRo^{-10/3}$. The numerical simulations also find that the heat flux scales as $c_pT\delta\rho\,v_\parallel\propto Ro^3$ such that we obtain $\delta\rho\propto Ro^{4/3}$. Inserting $\delta\rho\propto Ro^{4/3}$, $v_\perp\propto Ro^{7/3}$ and $\Omega\propto Ro^{-1}$, we can validate the force balance $\delta\rho\,g\approx\rho v_\perp\Omega$ in respect of $Ro$ scaling law, which indicates that this force balance is reliable. We now estimate magnetic energy $\langle B^2/\mu\rangle\approx\langle\delta\rho\,gl_\parallel\rangle\approx\langle\rho v_\perp\Omega l_\parallel\rangle$. Here we use the eddy length $l_\parallel$ elongated along rotation axis as the lengthescale of magnetic induction to estimate magnetic energy, i.e. $l_\parallel$ instead of $l$ in \eqref{l_B} and \eqref{magnetic-buoyancy}, because magnetic induction tends to take effect on the larger lengthscale $l_\parallel$ ($\approx lRo^{-10/3}$). Inserting the scaling $\delta\rho\propto Ro^{4/3}$ (or $v_\perp\propto Ro^{7/3}$ and $\Omega\propto Ro^{-1}$) and $l_\parallel\propto Ro^{-10/3}$, we immediately obtain the estimation for magnetic energy with fast rotation
\begin{equation}\label{B-fast}
\langle\frac{B^2}{\mu}\rangle\approx\langle\rho^{1/3}F^{2/3}\rangle Ro^{-2}.
\end{equation}
With fast rotation, magnetic energy depends on rotation rate and faster rotation (smaller $Ro$) corresponds to stronger magnetic field.

\section{Comparison with observations}

As stated in Section \S\ref{sec:motivation}, the fraction of X ray luminosity $L_X/L_{bol}$ represents the strength of magnetic field. The observations clearly show that faster rotation indeed corresponds to stronger $L_X/L_{bol}$ and the scaling law is $L_X/L_{bol}\propto Ro^{-2}$ \citep{wright2011}. On the other hand, the observations with different techniques give different relation of X ray luminosity and surface field, i.e. $L_X/L_{bol}\propto B_{surf}^{1.61}$ or $L_X/L_{bol}\propto B_{surf}^{2.25}$ \citep{vidotto2014}. Surface field $B_{surf}$ is  proportional to volume-averaged internal field $\langle B\rangle$, such that the observations yield the scaling law of internal magnetic field $\langle B\rangle\propto Ro^{-1.24}$ or $\langle B\rangle\propto Ro^{-0.89}$. Our prediction \eqref{B-fast} yields $\langle B\rangle\propto Ro^{-1}$, which is in good agreement with the observations.

The observations show that $L_X/L_{bol}$, or equivalently magnetic field, saturates at sufficiently low $Ro<0.01$ (it should be noted that $Ro=P_{rot}/\tau$ used for the observations differs from our definition $Ro=v/\Omega l$ by a factor $2\pi$). The saturation mechanism might be tentatively interpreted in this way: $l_\parallel$ grows to its maximum, namely the depth of stellar convection zone, at which buoyancy energy saturates, and consequently magnetic energy saturates. According to the scaling law in the last section we derived, the aspect ratio of columns is proportional to $Ro^{-10/3}$, so that when $Ro$ reaches $\sim 0.01$ the column is very thin with its aspect ratio $\sim 10^6$, i.e. column height (stellar convection zone depth) $\sim 10^5$ km and column width less than 1 km. These thin columns seem to be unstable but fast rotation can stabilize them \citep{chandrasekhar1961}.

\section*{Acknowledgments}
I discussed with Cai Tao about the numerical simulations of rotating turbulent convection. This work is supported by National Natural Science Foundation of China (11872246, 12041301) and Beijing Natural Science Foundation (1202015).

\bibliographystyle{apj}
\bibliography{paper}

\end{document}